\documentclass[pra,showpacs,superscriptaddress,eqsecnum]{revtex4}

\usepackage{bm}
\usepackage{graphicx}
\usepackage{amssymb}
\usepackage{amsmath}
\usepackage{url}

\def\tr{{\rm tr}}
\def\ket#1{|#1\rangle}
\def\bra#1{\langle#1|}

\def\rhoout{\rho_{\rm out}}
\def\rhoav{\overline{\rho\vphantom{\raisebox{0.02ex}{$\rho$}}}_{\rm out}}
\def\rhoavsub{\overline{\rho}_{\rm out}}

\def\gsim{\gtrsim}
\def\Fb{\,\overline{\vphantom{\raisebox{0.08ex}{F}}F}\!}
\def\Fbent{\,\overline{\vphantom{\raisebox{0.08ex}{F}}F}\!^{\hspace{0.25em}{\rm ent}}}
\def\dtwo{d^{\hspace{0.075em}2}\hspace{-0.055em}}

\begin{document}

\title{Teleportation fidelity as a probe of sub-Planck phase-space structure}

\author{A. J. Scott}
\email{andrew.scott@griffith.edu.au}
\affiliation{Centre for Quantum Computer Technology and Centre for Quantum Dynamics,
Griffith University, Brisbane, Queensland 4111, Australia}

\author{Carlton M. Caves}
\email{caves@info.phys.unm.edu}
\affiliation{Department of Physics and Astronomy, MSC07--4220, University of New Mexico, Albuquerque, NM 87131-0001, USA}
\affiliation{Department of Physics, University of Queensland, Brisbane, Queensland 4072, Australia}

\begin{abstract}
We investigate the connection between sub-Planck structure in the
Wigner function and the output fidelity of continuous-variable
teleportation protocols.  When the teleporting parties share a
two-mode squeezed state as an entangled resource, high fidelity in
the output state requires a squeezing large enough that the smallest
sub-Planck structures in an input pure state are teleported
faithfully.  We formulate this relationship, which leads to an
explicit relation between the fine-scale structure in the Wigner
function and large-scale extent of the Wigner function, and we treat
specific examples, including coherent, number, and random states and
states produced by chaotic dynamics.  We generalize the pure-state
results to teleportation of mixed states.

\end{abstract}

\maketitle

\section{Introduction}
\label{sec1}

Quantum
teleportation~\cite{Bennett1993,Vaidman1994,Braunstein1998,Braunstein2000}
transfers a quantum state from a quantum system owned by Alice to a
distant system owned by Bob at a cost, per qubit teleported, of one
bit of entanglement shared between Alice and Bob and two bits of
classical communication.  Bennett {\it et al.}~\cite{Bennett1993}
showed that a maximally entangled state shared between Alice and Bob
allows this process, which might na{\"\i}vely be presumed impossible
under quantum mechanical law: an {\em arbitrary\/} quantum state of a
system possessed by Alice can be perfectly transferred to a system
possessed by Bob through local operations and classical communication
alone. This apparent violation of the no-cloning
theorem~\cite{Wootters82,Dieks82} is explained by the complete
corruption of Alice's system after teleportation.

Teleportation has become a fundamental primitive for quantum
information processing~\cite{Nielsen00}.  Although the teleportation
protocol devised by Bennett {\it et al.} applies only to finite
dimensional systems (e.g., a spin-$\frac{1}{2}$ particle), it was
later extended to include continuous-variable systems (e.g., the
modes of an electromagnetic field) by Vaidman~\cite{Vaidman1994} and
then by Braunstein and Kimble~\cite{Braunstein1998}.  In the latter
case, a highly squeezed two-mode vacuum state is chosen for the
entangled resource, and only in the limit of infinite squeezing is
the teleported state a perfect replica of the original.  An
appropriate measure of the ``quality'' of the teleported state is its
probability overlap with the original, which is called the fidelity.

Experimental
demonstrations~\cite{Furusawa1998,Zhang2003,Bowen2003,Yonezawa2004,Takei2005a,Takei2005b,Yonezawa2007}
of continuous-variable teleportation have now achieved fidelities up
to $0.76\pm0.02$~\cite{Yonezawa2007} when the teleported state is
Gaussian.  It has been argued that teleportation of non-Gaussian
states with fidelities above 2/3 is necessary before these
experiments no longer afford interpretations formulated purely in
terms of classical correlations and thus require a quantum-mechanical
explanation~\cite{Caves2004}.  In the present article we show that
achieving such high fidelities in teleporting non-Gaussian states
requires faithful reproduction of the smallest sub-Planck structures
in the teleported state's Wigner function.

The existence and importance of sub-Planck structure in the Wigner
function was pointed out by Zurek~\cite{Zurek2001}.  He showed that
environmental perturbations to a quantum system at the sub-Planck
action scale, $a\sim 1/A$, where $A$ is the areal extent of the
phase-space region over which the system state has nonnegligible
Wigner function, are enough to cause orthogonality between perturbed
and unperturbed states and, hence, to drive decoherence.  This same
sensitivity to perturbation has been suggested as a way to improve
sensitivity in weak-force
detection~\cite{Munro2002,Toscano2006,Dalvit2006}, in much the same
way that the sub-Planck variation in a squeezed state along one
phase-space dimension can be used for this purpose.  Structures
analogous to the sub-Planck structures in the Wigner function of
non-Gaussian states have been identified in the classical
time-frequency domain of an electromagnetic field
mode~\cite{Praxmeyer2007}, and non-Gaussian states of an optical
field mode with sub-Planck structure have been generated and
investigated by Ourjoumtsev {\it et
al.}~\cite{Ourjoumtsev2006,Ourjoumtsev2007}.  We complement all this
previous work by demonstrating the importance of sub-Planck structure
to achieving high-fidelity continuous-variable teleportation.

We show that when the teleporting parties share a two-mode squeezed
state as an entangled resource, high fidelities in teleporting a pure
state $\rho$ require squeezing large enough that the smallest
sub-Planck structures in the Wigner function of $\rho$ are teleported
faithfully.  While this connection is reasonable on its face, we make
it mathematically explicit by showing that the rate of decrease in
fidelity is directly related to a natural measure of the fine-scale
structure in the Wigner function and, reciprocally, to a sensible
measure of the large-scale extent of the Wigner function, i.e., the
size of the region that encompasses all of the nonnegligible support
of the Wigner function.  This explicit connection takes the form
\begin{equation}\label{introeq1}
\left.\frac{d\Fb_{\!\rho}}{dt}\right|_{t=0} \;=\;
-\frac{1}{2}\left[(\Delta x)^2+(\Delta p)^2\right] \;=\;
-\frac{\displaystyle{\int dx\,dp\,\left|\nabla W'_{\!\rho}(x,p)\right|^2}}{\displaystyle{4\int dx\,dp\,W_{\!\rho}^{\prime\,2}(x,p)}}\;=\;
-\frac{\pi}{2}\int dx\,dp\,|\nabla W'_{\!\rho}(x,p)|^2 \;.
\end{equation}
In this equation $\Fb_{\!\rho}(t)$ is the average fidelity of the
teleported state for pure input state $\rho$, as a function of a
squeezing parameter $t=2e^{-2r}$, where $r$ is the standard squeezing
parameter (i.e., $t$ is the twice the ratio of the uncertainty in a
squeezed quadrature component to the uncertainty in an unsqueezed
quadrature); $(d\Fb_{\!\rho}/dt)_{t=0}$ is the rate of decrease of
fidelity away from the perfect fidelity achievable for infinite
squeezing ($t=0$); $(\Delta x)^2=\langle x^2\rangle-\langle
x\rangle^2$ and $(\Delta p)^2=\langle p^2\rangle-\langle p\rangle^2$
are the variances of $x$ and $p\,$; and $W'_\rho(x,p)$ is the Wigner
function of $\rho$, normalized to unity with respect to the
phase-space measure $dx\,dp$.  The first equality relates
$(d\Fb_{\!\rho}/dt)_{t=0}$ to a sensible measure of the extent of the
Wigner function in phase space, and the second equality relates it to
a natural measure of the fine-scale structure of the Wigner function.
The third equality follows from the fact that the obvious measure of
the (inverse) area of support of a Wigner function has the same value
for all pure states,
\begin{equation}\label{introeq2}
\int dx\,dp\,W_{\!\rho}^{\prime\,2}(x,p)\;=\;\frac{1}{2\pi}\;,
\end{equation}
corresponding to one Planck unit of action.  That the area of support
of the Wigner function is one Planck unit for all pure states already
tells one that when the large-scale extent is much bigger than a
Planck unit, the support must be very patchy within that extent,
leading to fine-scale structure.

The explicit connection among teleportation fidelity for pure states,
the fine-scale structure of the corresponding Wigner function, and
the large-scale extent of the Wigner function, all expressed in
Eq.~(\ref{introeq1}), is the key result of this paper. This
connection can be generalized to mixed states by using entanglement
fidelity~\cite{Schumacher1996} as the measure of success.  With
entanglement fidelity in place of fidelity, the first equality in
Eq.~(\ref{introeq1}) still holds, but the second and third equalities
do not.  The third equality can be replaced by a more general
measure, which expresses not the fine-scale structure of the mixed
state being teleported, but rather the fine-scale structure in any
purification of that mixed state. Indeed, the justification for the
generalization to mixed states is to let teleportation fidelity
identify for us an appropriate measure of the fine-scale structure
underlying a mixed state.

The article is organized as follows.  Section~\ref{secextra} reviews
the definitions of ordered characteristic functions and the
corresonding quasiprobability distributions.  The general procedure
for continuous-variable teleportation and its analysis in terms of
Wigner functions is outlined in Sec.~\ref{sec2A}.  The general
analysis is then specialized to squeezed-state teleportation in
Sec.~\ref{sec2B}, and the high-fidelity limit and the requirements on
the classical communication are discussed in Sec.~\ref{sec2C}.  In
Sec.~\ref{sec3} we explore the connection between teleportation
fidelity and sub-Planck structure.  In Sec.~\ref{sec4}, we consider
specific examples, including teleportation of coherent, number, and
random states. Sec.~\ref{sec5} generalizes our results to
teleportation of mixed states, and we conclude with a brief summary
in Sec.~\ref{sec6}.

\section{Characteristic functions and quasiprobability distributions}
\label{secextra}

Before proceeding to our analysis of teleportation, we summarize some
standard definitions used throughout the article---and our particular
conventions for those definitions~\cite{cavesnotes}.  Consider a mode
with annihilation operator $a=(x+ip)/\sqrt2$, where $x$ and $p$ can
be thought of as position and momentum operators and are sometimes
called quadrature components.  With this choice, we adopt natural
oscillator units in which position and momentum have the same
(dimensionless) units, scaled so that the vacuum uncertainties in
position and momentum are $\Delta x=\Delta p=1/\sqrt2$.

We use the displacement operator
\begin{equation}
D(a,\alpha)\;\equiv\;\exp(\alpha a^\dag-\alpha^*a)\;,
\end{equation}
written as a function of two variables, the first an operator and the
second a complex amplitude $\alpha=(\alpha_1+i\alpha)/\sqrt2$. The
real quantities $\alpha_1$ and $\alpha_2$ can be thought of as
dimensionless c-number variables for position and momentum,
respectively.  We use this decomposition for all Greek variables. The
displacement operator generates a coherent state with complex
amplitude $\alpha$ from the vacuum state, i.e.,
$D(a,\alpha)|0\rangle=|\alpha\rangle$.

When the operator slot in the displacement operator is filled by a
c-number, we get the two-dimensional Fourier expansion function,
\begin{equation}
D(\beta,\alpha)\;=\;e^{\alpha\beta^*-\alpha^*\beta}\;=\;e^{i(\alpha_2\beta_1-\alpha_1\beta_2)}\;,
\end{equation}
which satisfies
\begin{equation}
\int\frac{\dtwo\alpha}{\pi}D(\beta,\alpha)=\pi\delta(\beta)\;.
\end{equation}

The {\em $s$-ordered characteristic function\/} for the state $\rho$
is defined as
\begin{equation}
\Phi_\rho^{(s)}(\beta) \;\equiv\; e^{s|\beta|^2/2}\tr\big[\rho D(a,\beta)\big] \;.
\end{equation}
For $s=0$, $\Phi_\rho(\alpha)\equiv\Phi_\rho^{(0)}(\alpha)$ is the
{\em symmetrically ordered characteristic function\/}. The
(real-valued) {\em $s$-ordered quasiprobability distribution\/} is
$1/\pi$ times the Fourier transform of the corresponding
characteristic function,
\begin{equation}
W_\rho^{(s)}(\alpha) \;\equiv\;
\int\frac{\dtwo\beta}{\pi^2}\Phi_\rho^{(s)}(\beta)D(\beta,\alpha)
\;=\; \frac{1}{\pi}\tr\big[\rho\tilde D^{(s)}(a,\alpha)\big] \;.
\end{equation}
Here the integration measure is
$\dtwo\beta=d(\mathop{\text{Re}}\beta)\,d(\mathop{\text{Im}}\beta)=d\beta_1\,
d\beta_2/2$, and
\begin{equation}
\tilde D^{(s)}(a,\alpha)\;\equiv\;
\int\frac{\dtwo\beta}{\pi}D^{(s)}(a,\beta)D(\beta,\alpha)
\label{tildeD}
\end{equation}
is the (Hermitian) Fourier transform of the $s$-ordered displacement
operator $D^{(s)}(a,\beta)\equiv e^{s|\beta|^2/2}D(a,\beta)$.  These
functions satisfy
\begin{align}
\int \dtwo\alpha\, W_\rho^{(s)}(\alpha) &\;=\; \Phi_\rho^{(s)}(0) \;=\; \tr\rho \;=\; 1 \;, \\
\pi\!\int \dtwo\alpha\, W_{\rho_1}^{(-s)}(\alpha)W_{\rho_2}^{(s)}(\alpha) &\;=\; \int\frac{\dtwo\beta}{\pi}\Phi_{\rho_1}^{(-s)*}(\beta)\Phi_{\rho_2}^{(s)}(\beta) \;=\; \tr(\rho_1\rho_2) \;.
\end{align}

When $s=-1$ and $s=0$, we have
\begin{align}
&\tilde D^{(-1)}(a,\alpha)=|\alpha\rangle\langle\alpha|\;,\\
&\tilde D^{(0)}(a,\alpha)=2D(a,\alpha)(-1)^{a^\dagger a}D^\dagger(a,\alpha)=
2\int\frac{\dtwo\beta}{\pi}|\alpha+\beta\rangle\langle\alpha-\beta|D(\beta,\alpha)\;.
\end{align}
These allow us to write expressions for the $s=0$ and $s=-1$
quasidistributions in terms of the coherent-state matrix elements of
the density operator.  These two quasidistributions are called,
respectively, the {\em Wigner function\/} and the {\em Husimi\/} (or
{\em Q\/}) {\em function\/}:
\begin{align}
W_\rho(\alpha) &\;\equiv\; W_\rho^{(0)}(\alpha) \;=\; \frac{2}{\pi^2}\int \dtwo\beta\,
\bra{\alpha-\beta}\rho\ket{\alpha+\beta}D(\beta,\alpha) \;,\\
Q_\rho(\alpha) &\;\equiv\; W_\rho^{(-1)}(\alpha) \;=\; \frac{1}{\pi}\bra{\alpha}\rho\ket{\alpha} \;.
\end{align}
These expressions imply that $-{2}/{\pi}\leq
W_\rho(\alpha)\leq{2}/{\pi}$ and $0\leq Q_\rho(\alpha)\leq{1}/{\pi}$.
Since the Husimi function is nonnegative, it is a probability
distribution, rather than just a quasidistribution, albeit one that
cannot be too highly peaked.  The ordered characteristic functions
and corresponding quasidistributions were first explored
systematically by Cahill and Glauber~\cite{Cahill1969}.

When the Wigner function is written as a function of $x=\alpha_1$ and
$p=\alpha_2$, it is conventional to rescale it by a factor of two,
i.e., $W'_{\rho}(\alpha_1,\alpha_2)=W_\rho(\alpha)/2$, so that it is
normalized to unity with respect to $d\alpha_1\,d\alpha_2$, instead
of $\dtwo\alpha=d\alpha_1\,d\alpha_2/2$.  This rescaled Wigner
function appears in Eqs.~(\ref{introeq1}) and (\ref{introeq2}) and
the surrounding discussion, but we make no further use of it in this
paper.

\section{Continuous-variable teleportation}
\label{sec2}

\subsection{General procedure and analysis}
\label{sec2A}

Teleportation for continuous-variable systems~\cite{Braunstein1998}
is initiated when two parties, Alice and Bob, acquire two modes $A$
and $B$, prepared in the joint quantum state $\rho_{AB}$, with joint
Wigner function
\begin{equation}
W_{\rho_{AB}}(\alpha,\beta) \;=\;
\int\frac{\dtwo\mu}{\pi^2}\frac{\dtwo\gamma}{\pi^2}
\tr\big[\rho_{AB}D(a,\mu)\otimes
D(b,\gamma)\big]D(\mu,\alpha)D(\gamma,\beta)\;.
\label{WAB}
\end{equation}
Here Alice's mode is described by the annihilation operator
$a=(x_A+ip_A)/\sqrt2$ and Bob's by $b=(x_B+ip_B)/\sqrt2$.  We let
$\alpha$ and $\beta$ denote the respective complex-number variables
for the modes.  High-fidelity teleportation depends critically on the
entanglement contained in $\rho_{AB}$.  Thus the preparation of
$\rho_{AB}$ is generally done at some central point, after which the
two modes are distributed to Alice and Bob.

The state to be teleported is held by a third party called Victor. He
brings up to Alice a mode $V$ with annihilation operator
$v=(x_V+ip_V)/\sqrt2$; the corresponding complex amplitude is denoted
by $\nu$.  Victor's mode is prepared in a state $\rho$, with Wigner
function $W_\rho(\nu)$.  For the present, we allow $\rho$ to be pure
or mixed; we specialize to pure input states when we introduce the
teleportation fidelity below.  The overall Wigner function for the
three modes is $W_\rho(\nu)W_{\rho_{AB}}(\alpha,\beta)$.

Alice now measures the commuting variables $x_V+x_A$ and $p_V-p_A$,
or more succinctly, she measures the complex quantity
\begin{equation}
v+a^\dagger=\frac{1}{\sqrt2}(x_V+x_A)+\frac{i}{\sqrt2}(p_V-p_A)\;.
\end{equation}
The outcome, denoted by the complex number
$\xi=(\xi_1+i\xi_2)/\sqrt2$, occurs with probability
$p(\xi)\dtwo\xi$, where the probability density is
given by
\begin{equation}
p(\xi) \;=\; \int \dtwo\nu\,\dtwo\alpha\,\dtwo\beta\,
\delta(\nu+\alpha^*-\xi)W_\rho(\nu)W_{\rho_{AB}}(\alpha,\beta)
\;=\; \int \dtwo\nu\,W_\rho(\nu)W_{\rho_A}(\xi^*-\nu^*)
\label{pxi}
\end{equation}
and $\rho_A=\tr_B\rho_{AB}$ is the marginal state of Alice's mode.
The state of Bob's mode after the measurement, denoted by
$\rho'(\xi)$, is conditioned on result $\xi$ and has Wigner function
\begin{equation}
W_{\rho'}(\beta|\,\xi) \;=\; \frac{1}{p(\xi)}\int \dtwo\nu\,\dtwo\alpha\,
\delta(\nu+\alpha^*-\xi)W_\rho(\nu)W_{\rho_{AB}}(\alpha,\beta)
\;=\; \frac{1}{p(\xi)}\int \dtwo\nu\,
W_\rho(\nu)W_{\rho_{AB}}(\xi^*-\nu^*,\beta)\;.\label{telstate}
\end{equation}

Alice now sends the measurement result $\xi$ to Bob, who displaces
his mode by this amount; i.e., $x_B$ is displaced by $\xi_1$ and
$p_B$ is displaced by $\xi_2$, giving an output state $\rhoout(\xi)$
with Wigner function
\begin{equation}
W_{\rhoout}(\beta|\,\xi) \;=\;
W_{\rho'}(\beta-\xi|\,\xi) \;=\;
\frac{1}{p(\xi)}\int \dtwo\nu\,W_\rho(\nu)W_{\rho_{AB}}(\xi^*-\nu^*,\beta-\xi) \;.
\label{Wout}\end{equation}
The average of the output state over the possible measurement outcomes,
\begin{equation}
\rhoav \;=\; \int \dtwo\xi\,p(\xi)\rhoout(\xi)\;,
\label{avout}\end{equation}
has Wigner function
\begin{equation}
W_{\rhoavsub}(\beta) \;=\;
\int \dtwo\xi\,p(\xi)W_{\rhoout}(\beta|\,\xi)
\;=\; \int\dtwo\nu\,P(\nu)W_\rho(\beta-\nu)
\;=\; \int\dtwo\nu\,P(\beta-\nu)W_\rho(\nu)\;,\label{Wrhoav}
\end{equation}
where
\begin{equation}
P(\nu) \;\equiv\; \int \dtwo\alpha\,W_{\rho_{AB}}(\alpha,\nu-\alpha^*)
       \;=\; \int \dtwo\alpha\,\dtwo\beta\,\delta(\beta+\alpha^*-\nu)W_{\rho_{AB}}(\alpha,\beta) \label{measurenu1}
\end{equation}
is the probability density for obtaining result $\nu$ in a
measurement of the commuting observables $x_B+x_A$ and $p_B-p_A$,
i.e., in a measurement of
\begin{equation}
b+a^\dagger \;=\; \frac{1}{\sqrt2}(x_B+x_A)+\frac{i}{\sqrt2}(p_B-p_A)\;.\label{measurenu2}
\end{equation}

Since $W_\rho(\beta-\nu)$ is the Wigner function for the displaced
state $D(v,\nu)\rho D^\dagger(v,\nu)$, i.e.,
$W_\rho(\beta-\nu)=W_{D(v,\nu)\rho D^\dagger(v,\nu)}(\beta)$,
Eq.~(\ref{Wrhoav}) implies that the average output
state~(\ref{avout}) can be written as
\begin{equation}
\rhoav \;=\; \int \dtwo\nu\,P(\nu)D(b,\nu)\rho D^\dagger(b,\nu)\;,\label{avouttwo}
\end{equation}
where in this equation we regard the initial state $\rho$ as a state
of Bob's mode.  The average output state is an average over displaced
input states, the average controlled by the distribution $P(\nu)$. To
achieve high fidelity in teleporting $\rho$, no matter how it is
oriented in phase space, we need $P(\nu)$ to be a narrow distribution
in all directions in phase space, highly peaked at $\nu=0$.
Throughout this paper, what we mean by high-fidelity teleportation is
this ability faithfully to teleport $\rho$ and any rotation of
$\rho$.

The symmetrically ordered characteristic function for $\rhoav$ is
\begin{equation}
\Phi_{\rhoavsub}(\mu) \;\equiv\; \tr\bigl[\rhoav D(b,\mu)\bigr]
                   \;=\; \pi\tilde P(\mu)\Phi_\rho(\mu) \;,
\label{Phirhoav}
\end{equation}
where
\begin{equation}
\tilde P(\mu) \;\equiv\; \int \frac{\dtwo\nu}{\pi}\,P(\nu)D(\nu,\mu)
\end{equation}
is the Fourier transform of $P(\nu)$. Equation~(\ref{Phirhoav}) is
simply the Fourier transform of the corresponding Wigner-function
relation~(\ref{Wrhoav}).  We emphasize that $\tilde P(\mu)$ is not a
normalized probability distribution; rather its important properties
are that $\pi|\tilde P(\mu)|\le1$ and $\pi\tilde P(0)=1$.
Teleportation with high fidelity requires that $\pi\tilde P(\mu)$ be
close to 1 over the entire region for which $\Phi_\rho(\mu)$ is
nonnegligible.

Up to this point, we have not needed to say whether the input state
is pure or mixed, but we now want to measure the success of
teleportation in terms of the overlap of the input state with the
output state, called the fidelity.  For this purpose, we need to
assume that the input state $\rho=|\psi\rangle\langle\psi|$ is pure.
We maintain this assumption until Sec.~\ref{sec5}, where we
generalize our results to mixed states by using the entanglement
fidelity in place of the fidelity.  For outcome~$\xi$, the fidelity
of the output state with the input state is defined to be
$F_\rho(\xi)\equiv\langle\psi|\rhoout(\xi)|\psi\rangle$.  Thus the
average fidelity between input and output is
\begin{equation}
\Fb_\rho \;\equiv\; \int \dtwo\xi\,p(\xi)F_\rho(\xi) \;=\; \langle\psi|\,\rhoav|\psi\rangle\;.
\label{firstavF}
\end{equation}

We can manipulate the average fidelity into several forms, all of
which play a role in the subsequent discussion:
\begin{subequations}
\label{Fav}
\begin{align}
\Fb_\rho &\;=\; \int \dtwo\nu\,P(\nu)|\langle\psi|D(b,\nu)|\psi\rangle|^2 \;=\; \int \dtwo\nu\,P(\nu)|\Phi_\rho(\nu)|^2\label{Fav1}\\
         &\;=\; \pi\!\int \dtwo\beta\,\dtwo\nu\,\tilde P(\beta-\nu)W_\rho(\beta)W_\rho(\nu)\label{Fav2}\\
         &\;=\; \pi\!\int \dtwo\beta\,W_{\rhoavsub}(\beta)W_\rho(\beta) \;=\; \pi\!\int \dtwo\beta\,\dtwo\nu\,P(\beta-\nu)W_\rho(\beta)W_\rho(\nu)\label{Fav3}\\
         &\;=\; \int\frac{\dtwo\mu}{\pi}\, \Phi_{\rhoavsub}^*(\mu)\Phi_\rho(\mu) \;=\; \int \dtwo\mu\,\tilde P(\mu)|\Phi_\rho(\mu)|^2\label{Fav4}\;.
\end{align}
\end{subequations}
The first line, Eq.~(\ref{Fav1}), follows directly from inserting the
average output state~(\ref{avouttwo}) into the
expression~(\ref{firstavF}) for the average fidelity.  The second
line, Eq.~(\ref{Fav2}), comes from Fourier transforming the
quantities in the integrand of the first line.  The third line,
Eq.~(\ref{Fav3}), comes from rewriting $\bra{\psi}\rhoav\ket{\psi}$
as an overlap of the input and output Wigner functions,
$W_{\rhoavsub}(\beta)$ and $W_\rho(\beta)$, and then using
Eq.~(\ref{Wrhoav}) for $W_{\rhoavsub}(\beta)$.  Similarly, the fourth
line, Eq.~(\ref{Fav4}), comes from rewriting
$\bra{\psi}\rhoav\ket{\psi}$ as an overlap of the input and output
characteristic functions, $\Phi_{\rhoavsub}(\mu)$ and $\Phi_\rho(\mu)$,
and then using Eq.~(\ref{Phirhoav}) for $\Phi_{\rhoavsub}(\mu)$.  The
third and fourth lines are related to one another by a Fourier
transform of the quantities in the integrand.

These forms for the teleportation fidelity can be related to Zurek's
work on sub-Planck structures in phase space~\cite{Zurek2001}.  For a
given input state $\rho$, the Wigner function $W_\rho(\nu)$ has two
important length scales: (i)~a small scale $\ell$ over which the
Wigner function varies substantially and (ii)~a large scale $L$,
which is the extent of the region over which the Wigner function is
nonnegligible.  In the characteristic function $\Phi_\rho(\mu)$,
these scales appear inversely: (i)~a small scale $\pi/L$ over which,
in some phase-space direction(s), the characteristic function plunges
from 1 at $\mu=0$ to close to zero and (ii)~a large scale $\pi/\ell$,
which is the extent of the region over which the characteristic
function remains nonnegligible.  As pointed out by
Zurek~\cite{Zurek2001}, for pure states the large and small scales
are reciprocally related, i.e., $\ell L\sim1$.  We make this
reciprocal relationship explicit in Sec.~\ref{sec3}.

In Eqs.~(\ref{Fav}), the first and second lines form a pair under
Fourier transformation of the quantities in the integrand, and the
third and fourth lines constitute another such Fourier pair.  The
first and second lines relate the average fidelity to the large-scale
extent of the Wigner function $W_\rho(\nu)$, since good fidelity in
Eq.~(\ref{Fav2}) requires $\pi\tilde P$ to be close to 1 over the
entire extent of the Wigner function.  The third and fourth lines
relate the average fidelity to the fine-scale structure of the Wigner
function, since good fidelity in Eq.~(\ref{Fav3}) requires $P$ to be
narrow relative to the fine-scale structure in the Wigner function.
The strict connection between the two pairs and thus between the
large-scale and small-scale properties of the Wigner function comes
ultimately from the ability to write the average fidelity either as
an overlap or by using the expression~(\ref{avouttwo}) for the
average output state.

\subsection{Squeezed-state teleportation}
\label{sec2B}

We now specialize the above analysis to the case of squeezed-state
teleportation, where Alice and Bob choose a two-mode squeezed state
for $\rho_{AB}$.  We do a good job of teleporting when the
distribution $P(\nu)$ is narrow, i.e., when $x_A$ and $x_B$ are
tightly anti-correlated and $p_A$ and $p_B$ are tightly correlated.
Introducing the modes $c=(a+b)/\sqrt2=(x_C+ip_C)/\sqrt2$ and
$d=(a-b)/\sqrt2=(x_D+ip_D)/\sqrt2$, corresponding to c-number
variables $\gamma=(\gamma_1+i\gamma_2)/\sqrt2$ and
$\delta=(\delta_1+i\delta_2)/\sqrt2$, respectively, we see from
Eq.~(\ref{measurenu2}) that we want the variances of
$x_C=(x_A+x_B)/\sqrt2$ and $p_D=(p_A-p_B)/\sqrt2$ to be small.  Thus
a natural choice is to use squeezed vacuum states for modes $c$ and
$d$, with the variances of the quadrature components given by
\begin{align}
{\Delta x_C}^2 &\;=\; \frac{1}{2}\,e^{-2r}\;, \qquad {\Delta p_C}^2 \;=\; \frac{1}{2}\,e^{2r}\;,\nonumber\\
{\Delta x_D}^2 &\;=\; \frac{1}{2}\,e^{2r}\;, \qquad\;\; {\Delta p_D}^2 \;=\; \frac{1}{2}\,e^{-2r}\;,
\end{align}
where $r$ is the standard squeezing parameter. This state is a
two-mode squeezed vacuum state for modes $a$ and $b$ with Wigner
function
\begin{align}
W_{\rho_{CD}}(\gamma,\delta) &\;=\; \frac{4}{\pi^2}\exp\Bigl(-e^{2r}\gamma_1^2-e^{-2r}\gamma_2^2-e^{-2r}\delta_1^2-e^{2r}\delta_2^2\Bigr)\nonumber\\
                             &\;=\; \frac{4}{\pi^2}\exp\Bigl(-2|\beta+\alpha^*|^2/t-t|\beta-\alpha^*|^2/2\Bigr)
                             \;=\; W_{\rho_{AB}}(\alpha,\beta)\;, \label{squeezedWigner}
\end{align}
where we use the correspondence $\gamma=(\alpha+\beta)/\sqrt2$ and
$\delta=(\alpha-\beta)/\sqrt2$ and introduce a new squeezing
parameter $t\equiv 2e^{-2r}$.

Since $P(\nu)$ is the probability density to obtain result $\nu$ in a
measurement of $b+a^\dagger=x_C-ip_D$, the Wigner function
immediately implies that
\begin{equation}\label{Pnu}
       P(\nu) \;=\; \frac{2}{\pi t}\,e^{-2|\nu|^2/t}\;,
\end{equation}
with Fourier transform
\begin{equation}
\tilde P(\mu) \;=\; \frac{1}{\pi}\,e^{-t|\mu|^2/2}\;.
\label{tildeP}
\end{equation}
Notice that the two-mode Wigner function~(\ref{squeezedWigner}) can
be written as the product of these broad and narrow Gaussians, i.e.,
$W_{\rho_{AB}}(\alpha,\beta)=2tP(\beta+\alpha^*)\tilde
P(\beta-\alpha^*)$.  The marginal Wigner function for mode~$A$,
\begin{equation}\label{WA}
W_{\rho_A}(\alpha) \;=\;
\int \dtwo\beta\,W_{\rho_{AB}}(\alpha,\beta) \;=\; \frac{2t}{\pi(1+t^2/4)}\exp\left(-\frac{2t}{1+t^2/4}|\alpha|^2\right) \;\simeq\; 2t\tilde P(2\alpha)\;,
\end{equation}
is always broader than $\tilde P(2\alpha)$, but not by much for large
squeezing.  The final approximation holds in the limit of large
squeezing.

We can now specialize the important results of the preceding analysis
to the case of squeezed-state teleportation.  The average output
state~(\ref{avouttwo}) at Bob's end becomes
\begin{equation}\label{rhooutav}
\rhoav \;=\; \frac{2}{t}\int\frac{\dtwo\nu}{\pi}\,e^{-2|\nu|^2/t}D(b,\nu)\rho D^\dagger(b,\nu) \;.
\end{equation}
With $s\equiv-t$, the symmetrically ordered characteristic function
for $\rhoav$ becomes the $s$-ordered characteristic function for
$\rho$,
\begin{equation}
\Phi_{\rhoavsub}(\mu) \;=\; \pi\tilde P(\mu)\Phi_\rho(\mu) \;=\; e^{-t|\mu|^2/2}\Phi_\rho(\mu) \;=\; \Phi_\rho^{(s)}(\mu)\;,
\end{equation}
and hence the Wigner function for $\rhoav$ becomes the $s$-ordered
quasidistribution for $\rho$,
\begin{equation}
W_{\rhoavsub}(\beta)\;=\;W_\rho^{(s)}(\beta)
\;=\;\frac{2}{t}\int\frac{\dtwo\nu}{\pi}\,e^{-2|\beta-\nu|^2/t}W_\rho(\nu)\;.
\end{equation}

The various forms for the average fidelity in Eqs.~(\ref{Fav}) become
\begin{subequations}
\label{Fs}
\begin{align}
\Fb_\rho(t) &\;=\; \frac{2}{t}\int\frac{\dtwo\nu}{\pi}\,e^{-2|\nu|^2/t}|\Phi_\rho(\nu)|^2\label{Fs1}\\
                    &\;=\; \int \dtwo\beta\,\dtwo\nu\,\,e^{-t|\beta-\nu|^2/2}W_\rho(\beta)W_\rho(\nu)\;.\label{Fs2}\\
                    &\;=\; \pi\!\int \dtwo\beta\,W_\rho^{(s)}(\beta)W_\rho(\beta) \;=\; \frac{2}{t}\int \dtwo\beta\,\dtwo\nu\,e^{-2|\beta-\nu|^2/t}W_\rho(\beta)W_\rho(\nu)\label{Fs3}\\
                    &\;=\; \int\frac{\dtwo\mu}{\pi}\,\Phi_\rho^{(s)*}(\mu)\Phi_\rho(\mu) \;=\; \int\frac{\dtwo\mu}{\pi}\,e^{-t|\mu|^2/2}|\Phi_\rho(\mu)|^2\label{Fs4}
\end{align}
\end{subequations}
The first and fourth lines (or the second and third) show that the
teleportation fidelity obeys a scaling relation:
$t\Fb_\rho(t)/2=\Fb_\rho(4/t)$.  It is worth noting that these forms
of the squeezed-state teleportation fidelity show that it is
invariant under phase-space displacements and rotations of $\rho$.

The relevant range of $t$ is $0\le t\le 2$.  When $t=2$, Alice's and
Bob's modes are unentangled, each being in the vacuum state.  In this
case, the teleportation process reduces to a ``classical'' version of
teleportation.  Alice's joint measurement on her mode and Victor's
mode is a heterodyne measurement, for which the probability density
to obtain outcome $\xi$ is given by the Husimi function of $\rho$,
i.e., $p(\xi)=Q_\rho(\xi)$. When Bob receives the outcome $\xi$, he
displaces his vacuum state by $\xi$, yielding as output the coherent
state $\rhoout(\xi)=|\xi\rangle\langle\xi|$.  The resulting fidelity,
$F_\rho(\xi)=\langle\xi|\rho|\xi\rangle=\pi Q_\rho(\xi)$, is also
given by the Husimi function, yielding an average fidelity
\begin{equation}
\Fb_\rho(2)\;=\;\pi\!\int\dtwo\xi\,Q_\rho^2(\xi)\;=\;
\int\frac{\dtwo\mu}{\pi}|\Phi_\rho^{(-1)}(\mu)|^2\;=\;
\int\frac{\dtwo\mu}{\pi}\,e^{-|\mu|^2}|\Phi_\rho(\mu)|^2\;,
\end{equation}
in agreement with Eq.~(\ref{Fs4}).  Values $t>2$ are not unphysical,
but they do correspond to antisqueezing of the quadratures $x_C$ and
$p_D$ and thus to fidelities worse than this classical process.

\subsection{High-fidelity limit and communication requirements}
\label{sec2C}

In the limit of large squeezing, with $r$ going to infinity and $t$
going to zero, continuous-variable teleportation achieves high
fidelity.  The function $P(\nu)$ becomes a very narrow Gaussian,
approximating a delta function, while $2t\tilde P(\mu)$ becomes a
very broad Gaussian.  The average output state~(\ref{rhooutav}) is
obtained from the input state by Gaussian phase-space displacements
of characteristic size $\sqrt{t/2}=e^{-r}$.  The first form of the
average fidelity, Eq.~(\ref{Fs1}), shows that to get high fidelity
between input and output, $e^{-r}$ should be somewhat smaller than
the size $\pi/L$ of fine-scale structure in the characteristic
function of $\rho$; the fourth form, Eq.~(\ref{Fs4}), assures us that
the fidelity is near one when $e^r$ is somewhat larger than the
extent $\pi/\ell$ of the characteristic function.  Thus the first
and fourth forms express the reciprocal relationship between the
fine-scale structure and large-scale extent of a pure-state Wigner
function.

These same conclusions can be read off the Wigner-function forms of
the fidelity.  The second form of the average fidelity,
Eq.~(\ref{Fs2}), tells us that the fidelity is close to 1 when $e^r$
is somewhat larger than the extent $L$ of the Wigner function of
$\rho$, whereas the third form, Eq.~(\ref{Fs3}), assures us of high
fidelity when $e^{-r}$ is somewhat smaller than the scale $\ell$ of
fine-scale structure in the Wigner function.  High fidelity is
achieved when the available squeezing is sufficient to teleport
faithfully the fine-scale structure in the Wigner function.

In the high-fidelity limit, we can simplify the account of the
teleportation process.  To see what is going on, we take a closer
look at the probability density $p(\xi)$ for Alice to get outcome
$\xi$ in her measurement and at the conditional Wigner function
$W_{\rhoout}(\beta|\xi)$ of the teleported state.  The Wigner
function~(\ref{WA}) of Alice's marginal state is a broad Gaussian,
and thus the probability density $p(\xi)$ of Eq.~(\ref{pxi}) reduces
to the same broad Gaussian, displaced to account for the location of
$W_\rho(\nu)$, i.e.,
\begin{equation}
p(\xi)\;\simeq\; 2t\int \dtwo\nu\,W_\rho(\nu)\tilde
P(2\xi-2\nu)\;\simeq\; 2t\tilde P(2\xi-2\langle v\rangle) \;=\;
\frac{4e^{-2r}}{\pi}\exp\left(-4e^{-2r}|\xi-\langle v\rangle|^2\right) \;.
\label{approxp}
\end{equation}
In the second step, we take advantage of the fact that the broad
Gaussian $\tilde P$ is nearly constant over the extent of
$W_\rho(\nu)$ to evaluate $\nu$ in the broad Gaussian at a typical
point within the extent, taken here to be the mean value $\nu =
\langle v\rangle$ of Victor's mode.  The Wigner function~(\ref{Wout})
of the output state then becomes
\begin{align}
W_{\rhoout}(\beta|\,\xi)&\;=\;\frac{2t}{p(\xi)}\int\dtwo\nu\,W_\rho(\nu)P(\beta-\nu)\tilde P(\beta+\nu-2\xi)\nonumber\\
&\;\simeq\;\int \dtwo\nu\,W_\rho(\nu)P(\beta-\nu)\nonumber\\
&\;=\;W_{\rhoavsub}(\beta)\;=\;\frac{e^{2r}}{\pi}\int \dtwo\nu\,W_\rho(\nu)\exp\left(-e^{2r}|\beta-\nu|^2\right)\;.\label{approxWout}
\end{align}
In the second step we set $\beta=\nu = \langle v\rangle$ in the broad
Gaussian $\tilde P$ and use the above approximation for $p(\xi)$. The
result is that in the limit of high-fidelity teleportation, the
output state $\rhoout(\xi)$ is independent of the measurement result
$\xi$ and thus is the same as $\rhoav$.  In the limit
$t\rightarrow0$, $P(\beta-\nu)$ becomes a $\delta$ function, giving
perfect fidelity, i.e., $\rhoout(\xi)=\rhoav\rightarrow\rho$.

In the high-fidelity limit expressed by these equations, we can give
a a very simple Heisenberg-picture account of teleportation.  The
mean value of the measurement result $\xi$ is given by
$\langle\xi\rangle=\langle v\rangle$, with corresponding variances
${\Delta \xi_1}^2={\Delta \xi_2}^2=e^{2r}\!/4$. Following Alice's
measurement, we know that $v+a^\dag=\xi$, implying that $b=-a^\dag +
x_C -ip_D=v - \xi + x_C -ip_D$.  Displacing $b$ by $\xi$ then gives
$b=v + x_C -ip_D$.  Thus, after the teleportation process is
completed, Bob's mode is identical to Victor's initial mode, except
for contamination by the fluctuations in $x_C$ and $p_D$.  Since the
variances of $x_C$ and $p_D$ are both equal to $e^{-2r}/2$, the
convolution~(\ref{approxWout}) describes the output state
corresponding to this overall transformation of Bob's mode.

If Bob is to take full advantage of the squeezing resource, the value
of $\xi$ transmitted to him must allow him to perform displacements
with accuracy somewhat better than the uncertainties in $x_C$ and
$p_D$.  Thus the required number of bits in the transmission of
$\xi_1$ or $\xi_2$ must be roughly $\log(\Delta\xi_1/\Delta x_C)
=\log(e^{2r}/\sqrt2)$, making the total amount of information in
transmitting both $\xi_1$ and $\xi_2$ approximately equal to
$2\log(e^{2r}/\sqrt2)\simeq4r/\ln2$ bits.  The squeezing parameter
must be large enough to teleport the smallest phase-space structures
faithfully; i.e., $e^{-r}$ must be somewhat smaller than the
fine-scale structure in the Wigner function of $\rho$, of size
$\ell\sim1/L$. Thus Alice must transmit roughly $2\log L^2$ bits in
order to achieve high-fidelity teleportation.

Since $L^2$ is approximately the number of phase-space-local states
needed to represent $\rho$, the number of classical bits transmitted
has the standard form of 2 bits for each qubit of quantum
information.  Indeed, this argument gives us an independent way of
interpreting the required 2 bits of classical information per qubit.
The teleportation process must be able to distinguish $(L/\ell)^2$
phase-space regions to transmit all of the sub-Planck structure, and
this means transmitting $\log(L/\ell)^2\sim 2\log L^2$ bits of
classical information.

\section{Teleportation fidelity and sub-Planck structure}
\label{sec3}

The discussion of high-fidelity teleportation in Sec.~\ref{sec2C}
draws attention to how the expressions~(\ref{Fs}) connect
teleportation fidelity to the fine-scale structure and the
large-scale extent of the Wigner function of the state being
teleported.  In this section we make these connections explicit.

We begin by noting that the first two derivatives of the average
fidelity, as expressed in Eq.~(\ref{Fs4}),
\begin{gather}
\frac{d\Fb_\rho}{dt} \;=\; -\frac{1}{2}\int\frac{\dtwo\mu}{\pi}\,|\mu|^2e^{-t|\mu|^2/2}|\Phi_\rho(\mu)|^2 \;<\; 0\;, \label{dFdsone}\\
\frac{\dtwo\!\Fb_\rho}{dt^2} \;=\; \frac{1}{4}\int\frac{\dtwo\mu}{\pi}\,|\mu|^4e^{-t|\mu|^2/2}|\Phi_\rho(\mu)|^2 \;>\; 0\;,
\end{gather}
imply that $\Fb_\rho(t)$ is a strictly decreasing, strictly concave
function of $t$.  We can construct a characteristic scale for $t$ at
which teleportation becomes ineffective by approximating the average
fidelity for small $t$ as $\Fb_\rho(t)\simeq1+t(d\Fb_\rho/dt)_{t=0}$
and asking when this approximation goes to zero.  The result is a
critical value of $t$ given by
$t_c=\bigl|(d\Fb_\rho/dt)_{t=0}\bigr|^{-1}$. For $t\gsim t_c$,
teleportation becomes ineffective because the available squeezing is
unable to resolve the fine-scale phase-space structure of $\rho$.
Since $t=2e^{-2r}$, we convert this critical value into a phase-space
length that characterizes the size of the fine-scale structure by
defining
\begin{equation}
\ell_c\equiv\sqrt{t_c/2}\;=\;\frac{1}{\sqrt{2\bigl|(d\Fb_\rho/dt)_{t=0}\bigr|}}
\label{lcdef}\;.
\end{equation}

To evaluate $\ell_c$, we manipulate the derivative~(\ref{dFdsone}),
evaluated at $t=0$, through the following sequence of steps:
\begin{align}
\left.\frac{d\Fb_\rho}{dt}\right|_{t=0}
    &\;=\; -\frac{1}{2}\int\frac{\dtwo\mu}{\pi}\,|\mu|^2|\Phi_\rho(\mu)|^2\nonumber\\
    &\;=\; -\frac{1}{2}\int\frac{\dtwo\mu}{\pi}\,\dtwo\beta\,\delta(\beta-\mu)\,\mu^*\beta\,\Phi_\rho^*(\beta)\Phi_\rho(\mu)\nonumber\\
    &\;=\; -\frac{\pi}{2}\int \dtwo\nu\int\frac{\dtwo\beta}{\pi^2}\,\beta\,\Phi_\rho^*(\beta)D^*(\beta,\nu)\int\frac{\dtwo\mu}{\pi^2}\,\mu^*\,\Phi_\rho(\mu)D(\mu,\nu)\nonumber\\
    &\;=\; -\frac{\pi}{2}\int \dtwo\nu\,\left|\frac{\partial W_\rho(\nu)}{\partial\nu}\right|^2\;.
\label{Fderiv1a}
\end{align}
Noticing that
\begin{equation}
\frac{\partial W_\rho}{\partial\nu}
\;=\;\frac{1}{\sqrt2}\left(\frac{\partial W_\rho}{\partial\nu_1}-i\frac{\partial W_\rho}{\partial\nu_2}\right)\;,
\end{equation}
we can write
\begin{equation}
\left|\frac{\partial W_\rho}{\partial\nu}\right|^2
    \;=\; \frac{1}{2}\left[\left(\frac{\partial W_\rho}{\partial\nu_1}\right)^2+\left(\frac{\partial W_\rho}{\partial\nu_2}\right)^2\right]
    \;\equiv\; \frac{1}{2}\bigl|\nabla W_\rho(\nu_1,\nu_2)\bigr|^2\;.
    \label{gradW}
\end{equation}
Putting this together, we find that
\begin{equation}
\label{ellc}
\ell_c \;=\;\left(\frac{\pi}{4}\int d\nu_1\,d\nu_2\,\bigl|\nabla W_\rho(\nu_1,\nu_2)\bigr|^2\right)^{-1/2}\;.
\end{equation}
This measure is motivated here by the decrease of teleportation
fidelity, but it is a reasonable {\em a priori\/} measure of the
linear size of fine-scale structure in the Wigner function.  As noted
in the Introduction, for a general normalized distribution or
quasidistribution, $\ell_c$ quantifies the size of the fine-scale
structure multiplied by the square root of the area of support of the
distribution.  Since all pure-state Wigner functions have the same
area of support, corresponding to one Planck area, $\ell_c$ can be
interpreted as measuring of the size of the fine-scale structure.

We can also evaluate $\ell_c$ using the upper Fourier pair in
Eqs.~(\ref{Fs}), and this will relate our measure of fine-scale
structure to the large-scale extent of the Wigner function.
Differentiating and manipulating Eq.~(\ref{Fs2}), we have
\begin{align}
\left.\frac{d\Fb_\rho}{dt}\right|_{t=0}
    &\;=\; -\frac{1}{2}\int \dtwo\beta\,\dtwo\nu\,|\beta-\nu|^2W_\rho(\beta)W_\rho(\nu)\nonumber\\
    &\;=\; -\frac{1}{2}\int \dtwo\beta\,\dtwo\nu\,\bigl(|\beta|^2+|\nu|^2-\beta\nu^*-\beta^*\nu\bigr)W_\rho(\beta)W_\rho(\nu)\nonumber\\
    &\;=\; -\frac{1}{2}\left(\langle v^\dagger v\rangle+\langle vv^\dagger \rangle\right)+|\langle v \rangle|^2\nonumber\\
    &\;=\; -\frac{1}{2}\left({\Delta x_V}^2+{\Delta p_V}^2\right)\label{Fderiv1b}\;,
\end{align}
which gives
\begin{equation}
\ell_c\;=\;\left({\Delta x_V}^2+{\Delta p_V}^2\right)^{-1/2}\;=\;\frac{2}{L_c}\;,
\end{equation}
where we define
\begin{equation}
L_c\;\equiv\;2\sqrt{{\Delta x_V}^2+{\Delta p_V}^2}
\label{Lc}
\end{equation}
as a sensible measure of the linear extent of the region over which
the Wigner function is nonnegligible.  Through the use of the
subscript $c$, we emphasize the explicit definitions of $\ell_c$ and
$L_c$, as opposed to the somewhat loose use of $\ell$ and $L$ up
until now.  The relation
\begin{equation}\label{lL}
\ell_c L_c\;=\;2\;,
\end{equation}
arising from our considerations of teleportation fidelity, is a
rigorous expression of the reciprocal size of the fine-scale and
large-scale structures of a pure-state Wigner function.

It is easy to see that ${\Delta x_V}^2+{\Delta p_V}^2= (\Delta
x_V-\Delta p_V)^2+2\Delta x_V\Delta p_V\leq 2\Delta x_V\Delta p_V
\leq 1$, where the last step is the Heisenberg uncertainty principle.
Equality holds in both inequalities if and only if $\rho$ is a
coherent state.  Thus coherent states have the smallest initial rate
of decrease of average fidelity as $t$ increases from zero.
Equivalently, they have the smallest extent, $L_c=2$, and no
fine-scale structure, i.e., $\ell_c=1$.  That coherent states have
the smallest initial rate of decrease of fidelity is a reflection of
the fact that they have the highest teleportation fidelity for all
values of $t$, a fact we demonstrate in the next subsection.

\section{Example states}
\label{sec4}

We now examine various simple examples to develop our intuition for
the relationship between teleportation fidelity and sub-Planck
structure. In the following we investigate coherent, squeezed, and
number states, Zurek's compass state, a class of random states, and a
time-evolved state of a chaotic system.

\subsection{Coherent states}

All coherent states give the same average fidelity as the vacuum
state, for which
\begin{equation}
\Phi_{|0\rangle}(\mu) \;=\; \langle0|D(b,\mu)|0\rangle \;=\; e^{-|\mu|^2/2}\;.
\end{equation}
Therefore, the teleportation fidelity for any coherent state
$|\nu\rangle$ is
\begin{equation}\label{cohfidel}
\Fb_{|\nu\rangle}(t) \;=\; \int\frac{\dtwo\mu}{\pi}\,e^{-(1+t/2)|\mu|^2} \;=\; \frac{1}{1+t/2}\;,
\end{equation}
and $\ell_c=1$ ($L_c=2$).

Coherent states achieve the maximum teleportation fidelity for all
values of $t$, a fact we pause to demonstrate.  To do so, we notice
that the average fidelity in the form~(\ref{Fs2}) can be thought of
as the average value of $e^{-t|\beta-\nu|^2/2}$ with respect to a
product state $\rho\otimes\rho$ of two modes, $b$ and $v$, the joint
Wigner function for the two modes being
$W_{BV}(\beta,\nu)=W_\rho(\beta)W_\rho(\nu)$.  What we actually find
is the maximum value of this average value for all two-mode states
$\rho_{BV}$, not just tensor products of copies.

Using the fact that the Wigner function returns expectation values of
symmetrically ordered operator products, we can write the relevant
average as the expectation value of an operator $A_t$,
\begin{equation}
\int \dtwo\beta\,\dtwo\nu\,e^{-t|\beta-\nu|^2/2}W_{\rho_{BV}}(\beta,\nu)
\;=\; \tr(\rho_{BV}A_t)\;,
\end{equation}
where
\begin{equation}
A_t\;=\;\frac{1}{1+t/2}
\left(\frac{1-t/2}{1+t/2}\right)^{f^\dagger f}\;,
\label{At}
\end{equation}
with $f\equiv(b-v)/\sqrt2$.  The maximum of the expectation value is
given by the largest eigenvalue of $A_t$.  The eigenstates of $A_t$
are the number eigenstates of the mode with annihilation operator $f$
in tensor product with any state of the mode with annihilation
operator $g=(b+v)/\sqrt2$.  Since the factor in large parentheses in
the expression~(\ref{At}) has magnitude $\le1$, the largest
eigenvalue is $(1+t/2)^{-1}$, which occurs (uniquely when $t>0$) for
the vacuum state of mode $f$.  This establishes that the maximum
teleportation fidelity is the coherent-state
fidelity~(\ref{cohfidel}).

By returning to the case of interest, i.e., $\rho_{BV}$ being a
tensor product of two copies of a pure state, it is easy to show that
for $t>0$, the fidelity bound is saturated only by coherent states.

\begin{figure}[t]
\includegraphics[scale=1]{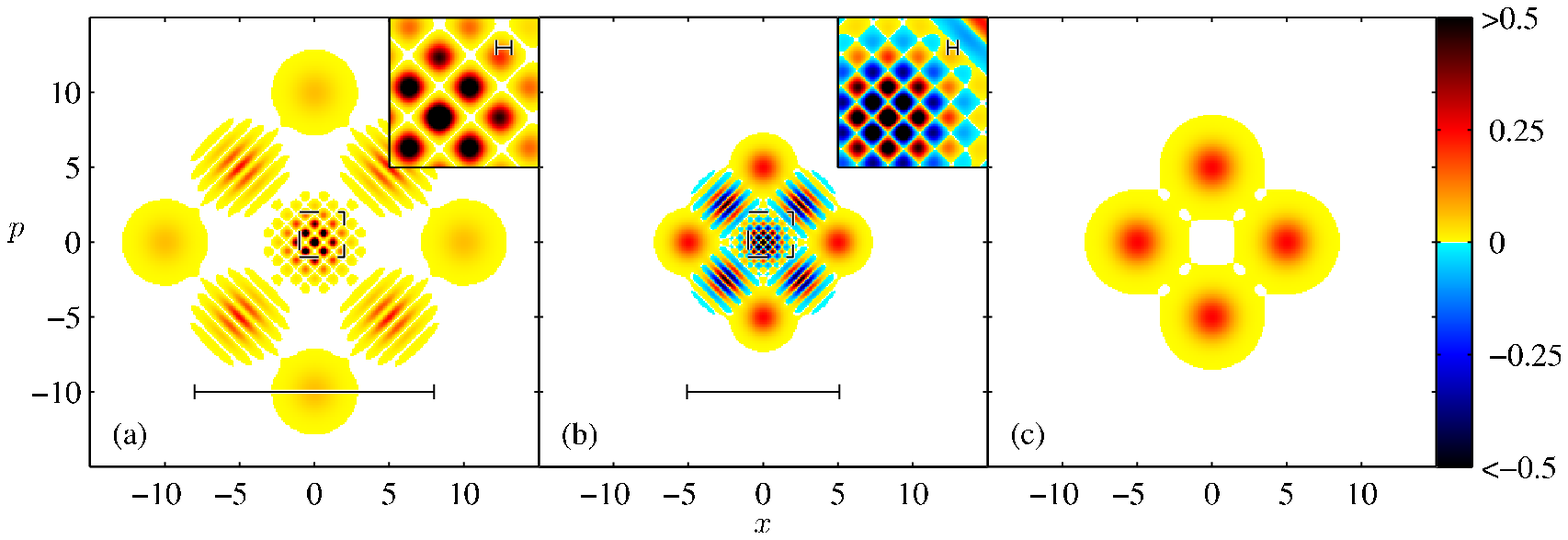}
\caption{Zurek's compass state with $a=5/\sqrt{2}$ as depicted by
(a)~$|\Phi_\rho|^2$, (b)~$(\pi/2)W_\rho$, and (c)~$\pi Q_\rho$.  The
Wigner function displays fine-scale structure on the scale
$\ell_c=1/\sqrt2 a=0.20$, and the size of its large-scale extent is
given roughly by $L_c=2/\ell_c=10$.  These scales are reversed in the
characteristic function, which has an extent characterized by
$\pi/\ell_c=\pi L_c/2=16$ and fine-scale structure on a scale
$\pi/L_c=\pi\ell_c/2=0.31$.  In~(a) and (b), the scale bars indicate
$\pi L_c/2$ and $L_c$, respectively, and the insets, which are
blow-ups of the regions bounded by dashed lines, show the fine-scale
structure in more detail, with scale bars indicating $\pi\ell_c/2$
and $\ell_c$, respectively.} \label{fig1}
\end{figure}
\begin{figure}[t]
\includegraphics[scale=1]{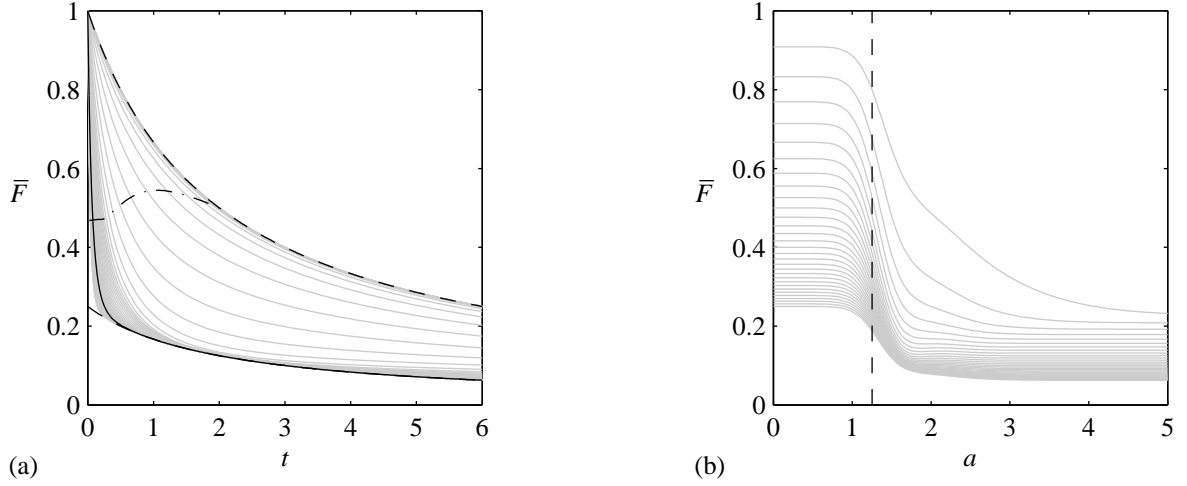}
\caption{(a)~Average fidelity~(\ref{zurekfidel}) of the compass
states for $a=0,0.1,0.2,\dots,5$ (lighter tones from top to bottom)
and of the compass state in Fig.~\ref{fig1} (full dark).  The upper
and lower bounds from Eq.~(\ref{zurekfidel}) are also shown (dashed),
as is the level of fidelity at the critical squeezing parameter
$t_c=2\ell_c^2$ (dash-dotted).  When $a=0$ the compass state is a
single coherent state centered at the origin with fidelity (upper
dashed curve) given by Eq.~(\ref{cohfidel}).  The
$a\rightarrow\infty$ bound $1/4(1+t/2)$ from Eq.~(\ref{zurekfidel})
is shown as the lower dashed curve, most of which is obscured by the
full dark line for the state of Fig.~\ref{fig1}. The dash-dotted
curve shows that to achieve a fidelity of approximately 1/2 or
better, we need to teleport with a squeezing parameter $t<t_c$.
(b)~Average fidelity of the compass states as a function of~$a$ for
$t=0.2,0.4,\dots,6$ (lighter tones from top to bottom).  The fidelity
declines sharply when $a=\sqrt{\pi/2}$ (dashed), at which point the
four coherent states are separated by a distance specified by a von
Neumann lattice.  This is the separation at which the interference
fringes and checkerboard pattern of Fig.~\ref{fig1} appear.} \label{fig2}
\end{figure}
\begin{figure}[t]
\includegraphics[scale=1]{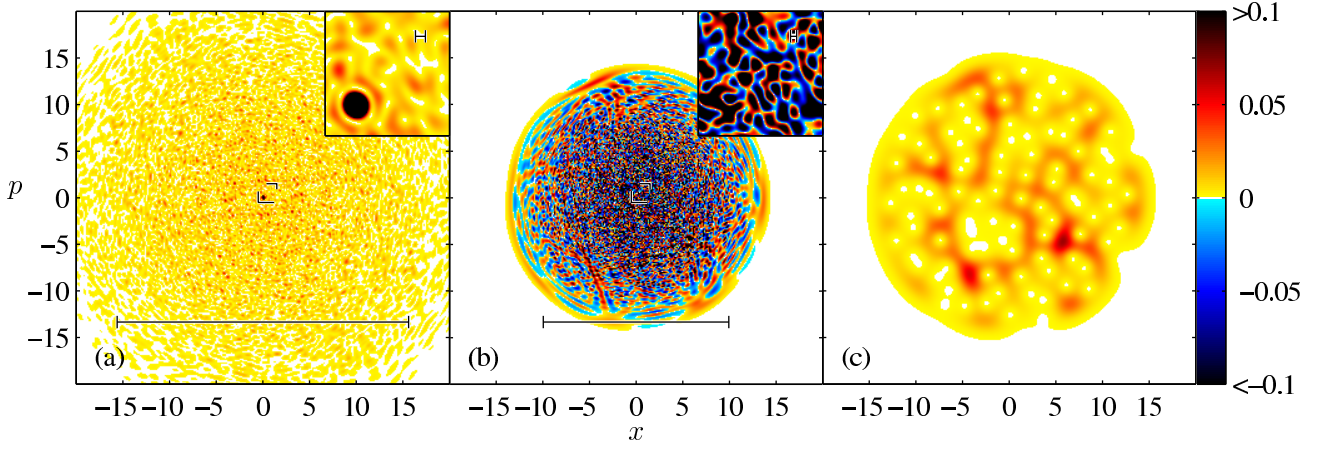}
\caption{A random state with $N=100$ as depicted by
(a)~$|\Phi_\rho|^2$, (b)~$(\pi/2)W_\rho$, and (c)~$\pi Q_\rho$.  The
Wigner function displays fine-scale structure on the scale
$\ell_c=1/\sqrt N=0.10$, and the size of its large-scale extent is
given approximately by $L_c=2/\ell_c=20$.  These scales are reversed
in the characteristic function, which has an extent characterized by
$\pi/\ell_c=\pi L_c/2=31$ and fine-scale structure on a scale
$\pi/L_c=\pi\ell_c/2=0.16$.  The scale bars and insets in~(a) and (b)
are as in Fig.~\ref{fig1}.} \label{fig3}
\end{figure}
\begin{figure}[t]
\includegraphics[scale=1]{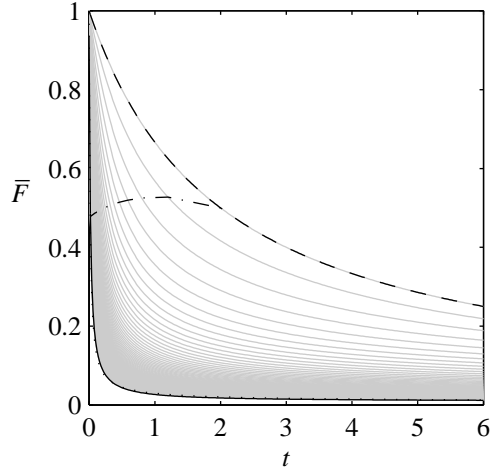}
\caption{The average fidelity~(\ref{randfidel}) for random states
with $N=1,2,\dots,100$ (lighter tones from top to bottom).  The case
$N=1$, which reduces to the coherent-state fidelity~(\ref{cohfidel}),
is highlighted (dashed), and the level of fidelity at the critical
squeezing value $t_c$ is plotted (dash-dotted), showing again that to
achieve a fidelity of approximately $1/2$ or better requires $t<t_c$.
The fidelities of the particular random state in Fig.~\ref{fig4}
(dark full) and the chaotic state in Fig.~\ref{fig5} (dotted) are
also drawn; they are essentially indistinguishable.} \label{fig4}
\end{figure}
\begin{figure}[t]
\includegraphics[scale=1]{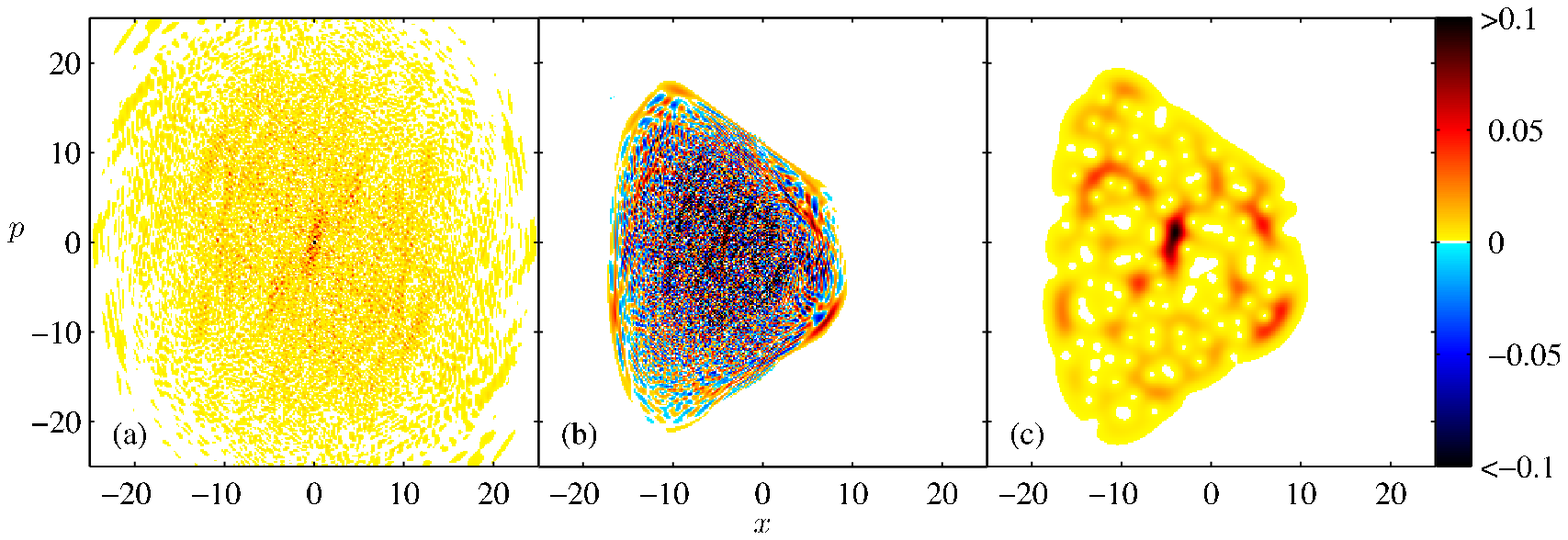}
\caption{A chaotic state produced by the driven double-well
Hamiltonian~(\ref{chaopot}) as depicted by (a)~$|\Phi_\rho|^2$,
(b)~$(\pi/2)W_\rho$, and (c)~$\pi Q_\rho$.} \label{fig5}
\end{figure}

\subsection{Squeezed states}
\label{sec4squeezed}

All squeezed states with the same squeezing parameter $u$ have the
same average fidelity; so we need only calculate the fidelity for a
squeezed vacuum state
\begin{equation}
\ket{0,u}\;\equiv\; e^{u(b^2-b^{\dag 2})/2}\ket{0}\;.
\end{equation}
Using
\begin{equation}
e^{-u(b^2-b^{\dag 2})/2}D(b,\nu)e^{u(b^2-b^{\dag 2})/2}
\;=\;D(b,\mu\cosh u+\mu^*\sinh u)\;=\;
D\big(b,(e^{u}\mu_1+ie^{-u}\mu_2)/\sqrt{2}\big)\;,
\end{equation}
we find the characteristic function,
\begin{equation}
\Phi_{|0,u\rangle}(\mu) \;=\; e^{-(e^{2u}\mu_1^2+e^{-2u}\mu_2^2)/4} \;.
\end{equation}
The teleportation fidelity for any squeezed state with squeezing
parameter $u$ is then
\begin{equation}
\Fb_{|\nu,u\rangle}(t)
    \;=\; \int\frac{\dtwo\mu}{\pi}\,e^{-t|\mu|^2/2}e^{-(e^{2u}\mu_1^2+e^{-2u}\mu_2^2)/2}
    \;=\; \frac{1}{\sqrt{1+t\cosh2u+t^2/4}}\;,
\end{equation}
giving $\ell_c=1/\sqrt{\cosh2u}$ ($L_c=2\sqrt{\cosh2u}$).  For a
highly squeezed state ($u\gg1$), we have $\ell_c\simeq\sqrt2 e^{-u}$
and $L_c\simeq \sqrt2 e^u$.

\subsection{Number states}

The number-state matrix elements of the displacement operator are
given by~\cite{Cahill1969}
\begin{equation}
\langle m|D(b,\mu)|n\rangle \;=\;
\begin{cases}
\sqrt{n!/m!}\,e^{-|\mu|^2/2}\mu^{m-n}L_n^{(m-n)}\bigl(|\mu|^2\bigr)\,, &\; \text{if $m\ge n$}\,, \\
\sqrt{m!/n!}\,e^{-|\mu|^2/2}(-\mu^*)^{n-m}L_m^{(n-m)}\bigl(|\mu|^2\bigr)\,, &\; \text{otherwise}\,,
\end{cases} \label{Dnumbasis}
\end{equation}
where $L_n^m(x)$ are generalized Laguerre
polynomials~\cite{Abramowitz}.  The characteristic function of a
number state is a diagonal matrix element:
\begin{equation}
\Phi_{|n\rangle}(\mu) \;=\; \langle n|D(b,\mu)|n\rangle
                      \;=\; e^{-|\mu|^2/2}L_n\bigl(|\mu|^2\bigr)\;.
\end{equation}
The number-state fidelity is thus
\begin{align}
\Fb_{|n\rangle}(t) &\;=\; \int\frac{\dtwo\mu}{\pi} e^{-(1+t/2)|\mu|^2}\big[L_n\bigl(|\mu|^2\bigr)\big]^2 \\
                   &\;=\; \int_0^\infty dx\; e^{-(1+t/2)x} \big[L_n(x)\big]^2 \\
                   &\;=\; \frac{(2n)!}{(n!)^2}(1+t/2)^{-2n-1}F\big(-n,-n;-2n;1-t^2/4\big) \\
                   &\;=\; \frac{(1-t/2)^n}{(1+t/2)^{n+1}}\,P_n\!\left(\frac{1+t^2/4}{1-t^2/4}\right)\;,
\end{align}
where $F(a,b;c;x)$ is a hypergeometric function, $P_n(x)$ is the
Legendre polynomial, and we have used formulae found in
Refs.~\cite{Abramowitz,Gradshtein}.  Using Eq.~(\ref{lcdef}), we find
that $\ell_c=1/\sqrt{2n+1}$.  More directly, we can use the
number-state variances $(\Delta x)^2=(\Delta p)^2=n+\frac{1}{2}$ to
find $2/\ell_c=L_c=2\sqrt{2n+1}$.

\subsection{Zurek's compass state}

Zurek introduced the ``compass state'' in his original article on
sub-Planck structures~\cite{Zurek2001}.  It is a superposition of
four coherent states at positions $(x,p)=(\sqrt{2}a,0)$,
$(-\sqrt{2}a,0)$, $(0,\sqrt{2}a)$, and $(0,-\sqrt{2}a)$:
\begin{equation}
\ket{\maltese} \;=\;
\frac{|a\rangle + |{-}a\rangle + |ia\rangle + |{-}ia\rangle}{2e^{-a^2/2}\sqrt{2(\cosh{a^2}+\cos{a^2})}}\;.
\end{equation}
A straightforward, but laborious calculation yields the average
teleportation fidelity:
\begin{align}
\Fb_{\ket{\maltese}}(t)
    &\;=\; \frac{1}{4(1+t/2)}\left[1+
    \frac{\displaystyle{\left(\cosh\frac{2-t}{2+t}a^2+\cos\frac{2-t}{2+t}a^2\right)^2
    +2\!\left(\cosh a^2+\cos\frac{2-t}{2+t}a^2\right)\!\!\left(\cos a^2+\cosh\frac{2-t}{2+t}a^2\right)}}
    {\displaystyle{\vphantom{\bigg(}\left(\cosh a^2+\cos a^2\right)^2}}\right] \nonumber \\
    &\;\rightarrow\;
    \begin{cases}
        \displaystyle{\vphantom{\Bigg(}\frac{1}{1+t/2}}\;,      & \text{for $a\rightarrow 0$}\,,  \\
        \displaystyle{\vphantom{\Bigg(}\frac{1}{4(1+t/2)}}\;,   & \text{for $a\rightarrow\infty$}\,.
    \end{cases}\label{zurekfidel}
\end{align}
Differentiating this expression gives
\begin{equation}
\ell_c \;=\;
\left(1+2a^2 \frac{\sinh a^2 - \sin a^2}{\cosh a^2 + \cos a^2}\right)^{-1/2}
\;\simeq\; \frac{1}{\sqrt{2}a} \quad \text{for $a\rightarrow\infty$}\;.
\end{equation}
More directly, one can evaluate $(\Delta x)^2+(\Delta p)^2$ and use
Eq.~(\ref{Fderiv1b}) to find $\ell_c=2/L_c$.

The Wigner function of a compass state with $a=5/\sqrt{2}$, which is
well into the regime of large $a$, is shown in Fig.~\ref{fig1}(b).
Interference fringes form between adjacent coherent states, combining
at the origin to create a checkerboard of fine structure at a scale
on the order of $\ell_c=1/\sqrt{2}a=1/5$.  Complementary behavior is
displayed in the Fourier transform, the characteristic function,
whose absolute square is plotted in Fig.~\ref{fig1}(a). The Husimi
function [Fig.~\ref{fig1}(c)] can be viewed as a Gaussian smoothing
of the Wigner function.  The average teleportation
fidelity~(\ref{zurekfidel}) for a compass state is plotted in
Fig.~\ref{fig2}(a) for various values of $a$.  The caption explains
in detail the various features plotted in Fig.~\ref{fig2}.

\subsection{Random states}

We now consider states of the form
\begin{equation}
\ket{\psi} \;=\; \sum_{n=0}^{N-1} c_n \ket{n}
\label{Nstate}
\end{equation}
where the states $\ket{n}$ are the number eigenstates and the
coefficients $c_n$ form a a random complex unit vector in $N$
dimensions under the uniform measure, i.e., a random point on the
unit sphere in $2N$ dimensions.  These states are conjectured to have
the same statistical properties as an eigenstate (or long-time
evolved state) of a chaotic system with a region of ergodicity within
a circle of radius $r=\sqrt{2N}$ in phase space~\cite{Leboeuf1999}.
Figure~\ref{fig3} plots the absolute square of the characteristic
function, the Wigner function, and the Husimi function for an example
random state with $N=100$.

To calculate the average fidelity of a random state, we use the
fidelity in the form~(\ref{Fs4}), noting that the absolute square of
the characteristic function for any state of the form~(\ref{Nstate})
is given by
\begin{equation}
|\Phi_\rho(\mu)|^2\;=\;
\left|\bra{\psi}D(b,\mu)\ket{\psi}\right|^2
\;=\; \sum_{m,n,k,j=0}^{N-1} c_n c_m^* c_j c_k^* \bra{m}D(b,\mu)\ket{n}\bra{k}D^\dag(b,\mu)\ket{j}\;,
\end{equation}
The necessary averages over random states as described above are
given by
\begin{equation}
\mathrm{E}\!\left[c_n c_m^* c_j c_k^*\right] \;=\;
\frac{\delta_{mn}\delta_{jk}+\delta_{mj}\delta_{nk}}{N(N+1)}\;. \label{avgc}
\end{equation}
These averages imply that $\mathrm{E}[c_n c_m^*]=\delta_{nm}/N$.  As
a result, we have
\begin{equation}
\mathrm{E}\!\left[|\Phi_\rho(\mu)|^2\right]\;=\;
\frac{1}{N(N+1)} \sum_{m,n=0}^{N-1} \Big( \bra{m}D(b,\mu)\ket{m}\bra{n}D^\dag(b,\mu)\ket{n}+\bra{m}D(b,\mu)\ket{n}\bra{n}D^\dag(b,\mu)\ket{m}\Big)\;.
\end{equation}
Using Eq.~(\ref{Dnumbasis}) and Ref.~\cite{Gradshtein}, one can show
that
\begin{align}
\int \frac{\dtwo\mu}{\pi} e^{-t|\mu|^2/2} \bra{m}D(b,\mu)\ket{n}\bra{n}D^\dag(b,\mu)\ket{m} &\;=\; \frac{(m+n)!}{m!\,n!\,(1+t/2)^{m+n+1}}F\big(-m,-n;-m-n;1-t^2/4\big)\;, \\
\int \frac{\dtwo\mu}{\pi} e^{-t|\mu|^2/2} \bra{m}D(b,\mu)\ket{m}\bra{n}D^\dag(b,\mu)\ket{n} &\;=\; \frac{(m+n)!(t/2)^{m+n}}{m!\,n!\,(1+t/2)^{m+n+1}}F\big(-m,-n;-m-n;1-4/t^2\big)\;,
\end{align}
which gives
\begin{multline}
\label{randfidel}
\mathrm{E}\!\left[\Fb_{\ket{\psi}}(t)\right] \;=\; \frac{1}{N(N+1)} \sum_{m,n=0}^{N-1} \frac{(m+n)!}{m!\,n!\,(1+t/2)^{m+n+1}}\Big[F\big(-m,-n;-m-n;1-t^2/4\big) \\
+(t/2)^{m+n}F\big(-m,-n;-m-n;1-4/t^2\big)\Big]\;.
\end{multline}
This average fidelity is plotted in Fig.~\ref{fig4} for
$N=1,2,\dots,100$ (lighter tones).

To calculate the slope of the fidelity at $t=0$, it is easiest to
work from Eq.~(\ref{Fderiv1b}) to obtain
\begin{equation}
-\frac{1}{2}\left[(\Delta x)^2+(\Delta p)^2\right]\;=\;
\left.\frac{d\Fb_{\ket{\psi}}}{dt}\right|_{t=0} \;=\;
-\frac{1}{2}\sum_{n=0}^{N-1}(2n+1)|c_n|^2+\sum_{m,n=0}^{N-2}\sqrt{(n+1)(m+1)}c_n^*c_{n+1}c_mc_{m+1}^*\;.
\end{equation}
Averaging over random states leads to
\begin{equation}
-\frac{1}{2}\mathrm{E}\!\left[(\Delta x)^2+(\Delta p)^2\right]
\;=\;\mathrm{E}\!\left[\left.\frac{d\Fb_{\ket{\psi}}}{dt}\right|_{t=0}\right]
\;=\; \left.\frac{d\,\mathrm{E}\bigl[\Fb_{\ket{\psi}}\bigr]}{dt}\right|_{t=0}
\;=\; -\,\frac{N^2+1}{2(N+1)}\;.
\end{equation}
We now define our measures of small- and large-scale structure to be
\begin{equation}
2/\ell_c\;=\;L_c\;=\;2\sqrt{\mathrm{E}\!\left[(\Delta x)^2+(\Delta p)^2\right]}
\;=\;2\sqrt{\frac{N^2+1}{N+1}}\;\simeq\;2\sqrt N \quad \text{for $N\rightarrow\infty$}\;,
\end{equation}
ignoring that the averaging over random states doesn't commute with
taking square roots and reciprocals, although it will do so very
closely when $N$ is large.

\subsection{Chaotic state}

We now consider briefly sub-Planck structure in states of a chaotic
Hamiltonian.  Consider a long-time evolved state of the driven
double-well potential, corresponding to Hamiltonian
\begin{equation}
H \;=\; 5p^2-8x^2+0.05x^4+65x\cos(2\pi t)\;.\label{chaopot}
\end{equation}
For the parameters chosen in this Hamiltonian, this system is chaotic
in the classical limit.  By choosing an initial coherent state
centered at $(x,p)=(-8,4)$ and then evolving the system from $t=0$ to
$t=5$, we obtain the chaotic state pictured in Fig.~\ref{fig5}.  The
Wigner function displays an abundant fine-scale structure that
qualitatively resembles that of the random state depicted in
Fig.~\ref{fig3}.  It is no surprise that the fidelity curve also
follows that for random states.  This is plotted in Fig.~\ref{fig4}
as the dotted curve, but is obscured behind the fidelity for the
random state (full dark).

\section{Mixed-state teleportation and entanglement fidelity}
\label{sec5}

In the preceding sections the teleported state was assumed to be
pure.  Since teleportation is a linear operation, the procedure
outlined in Sec.~\ref{sec2} works equally well for mixed states.  The
overlap between input and output states, however, is no longer an
appropriate measure of teleportation fidelity.  A suitable measure
for assessing the fidelity with which a mixed state $\rho$ is
teleported is the {\em entanglement
fidelity\/}~\cite{Schumacher1996}, which is the fidelity for
teleporting Victor's half of a purification of $\rho$, thus
transferring the entanglement to Bob (entanglement swapping). It is
quite easy to see how to generalize all of our results to
entanglement fidelity.  Victor's mode is now entangled with another
mode, labeled by $U$, which has annihilation operator $u$ and
corresponding complex variable $\mu$.  The joint state of $U$ and $V$
is a pure state $\rho_{UV}=|\psi_{UV}\rangle\langle\psi_{UV}|$, which
purifies Victor's state, i.e., ${\rm tr}_U(\rho_{UV})=\rho$.  The
results of Sec.~\ref{sec2} [Eqs.~(\ref{pxi}), (\ref{Wout}),
(\ref{Wrhoav}), (\ref{avouttwo}) and (\ref{Phirhoav})] generalize
immediately to the following:
\begin{align}
p(\xi) &\;=\; \int \dtwo\mu\,\dtwo\nu\,W_{\rho_{UV}}(\mu,\nu)W_{\rho_A}(\xi^*-\nu^*)\;, \\
W_{\rhoout}(\mu,\beta|\,\xi) &\;=\; \frac{1}{p(\xi)}\int \dtwo\nu\,W_{\rho_{UV}}(\mu,\nu)W_{\rho_{AB}}(\xi^*-\nu^*,\beta-\xi)\;, \\
W_{\rhoavsub}(\mu,\beta) &\;=\; \int \dtwo\nu\,P(\nu)W_{\rho_{UV}}(\mu,\beta-\nu)\;, \\
\rhoav &\;=\; \int \dtwo\nu\,P(\nu)\,\left[I\otimes D(b,\nu)\right]\,\rho_{UV}\left[I\otimes D^\dagger(b,\nu)\right]\;, \\
\Phi_{\rhoavsub}(\nu,\alpha) &\;=\; \pi\tilde P(\alpha)\Phi_{\rho_{UV}}(\nu,\alpha)\;,
\end{align}

The entanglement fidelity averaged over outcomes $\xi$ is
\begin{equation}\label{Fent}
\Fbent_\rho \;=\; \langle\psi_{UV}|\,\rhoav|\psi_{UV}\rangle
\;=\; \pi^2\int \dtwo\mu\,\dtwo\beta\,W_{\rhoavsub}(\mu,\beta)W_{\rho_{UV}}(\mu,\beta)\;;
\end{equation}
notice that when $\rho$ is pure, so that $|\psi_{UV}\rangle$ is a
product state, the entanglement fidelity reduces to the ordinary
fidelity~(\ref{firstavF}).  The average entanglement fidelity can be
put in the four forms analogous to Eqs.~(\ref{Fav}):
\begin{subequations}
\label{Fentone}
\begin{align}
\Fbent_\rho
    &\;=\; \int \dtwo\nu\,P(\nu)|\Phi_\rho(\nu)|^2\label{Fentone1}\\
    &\;=\; \pi\!\int \dtwo\beta\,\dtwo\nu\,\tilde P(\beta-\nu)W_\rho(\beta)W_\rho(\nu)\label{Fentone2}\\
    &\;=\; \pi\!\int \dtwo\beta\,\dtwo\nu\,P(\beta-\nu)\left[\pi\!\int \dtwo\mu\,W_{\rho_{UV}}(\mu,\beta)W_{\rho_{UV}}(\mu,\nu)\right]\;,\label{Fentone3}\\
    &\;=\; \int \dtwo\alpha\,\tilde P(\alpha)\left[\int\frac{\dtwo\nu}{\pi}\,|\Phi_{\rho_{UV}}(\nu,\alpha)|^2\right]\;.\label{Fentone4}
\end{align}
\end{subequations}
The first line, Eq.~(\ref{Fentone1}), comes from inserting $\rhoav$
into the first form of the fidelity in Eq.~(\ref{Fent}).  Fourier
transforming the terms in the integrand yields the second line,
Eq.~(\ref{Fentone2}).  This Fourier pair is identical to the upper
two lines in the pure-state teleportation fidelity~(\ref{Fav}); they
relate the entanglement fidelity to the large-scale extent of the
Wigner function of $\rho$ or, equivalently, to the fine-scale
structure of the characteristic function of $\rho$. The third and
fourth lines come from writing the entanglement fidelity as an
overlap of joint Wigner functions or joint characteristic functions.
These two lines also constitute a Fourier pair, which relates the
average entanglement fidelity to the fine-scale structure in the
joint Wigner function $W_{\rho_{UV}}(\mu,\beta)$ or, equivalently, to
the large-scale extent of the joint characteristic function
$\Phi_{\rho_{UV}}(\nu,\alpha)$. This Fourier pair is different from
the corresponding third and fourth lines for the pure-state
fidelity~(\ref{Fav}) precisely because these properties appear in the
joint functions, not in the Wigner and characteristic functions for
$\rho$ alone.

We can, however, convert the third and fourth lines to forms that
involve only system operators.  For this purpose, we focus first on
the characteristic-function integral
\begin{equation}
\int\frac{\dtwo\nu}{\pi}\,|\Phi_{\rho_{UV}}(\nu,\alpha)|^2 \;=\;
\int\frac{\dtwo\nu}{\pi}\,|\Phi_{\rho_{UV}}(-\nu^*,\alpha)|^2\;.
\label{jointPhi}
\end{equation}
The reason for the term on the right becomes clear as we manipulate
this integral below.  Since the entanglement fidelity is independent
of which purification is used, we can use the purification
\begin{equation}
|\psi_{UV}\rangle \;=\; \sum_{n=0}^\infty|n\rangle_U\otimes\sqrt\rho\,|n\rangle_V\;,
\end{equation}
where $|n\rangle_U$ and $|n\rangle_V$ denote number states for $U$
and $V$, respectively.

The characteristic function now becomes
\begin{align}
\Phi_{\rho_{UV}}(-\nu^*,\alpha)
    &\;=\; \langle\psi_{UV}|D(u,-\nu^*)\otimes D(v,\alpha)|\psi_{UV}\rangle\nonumber\\
    &\;=\; \sum_{m,n}{}_U\langle n|D(u,-\nu^*)|m\rangle_U{}_V\langle n|\sqrt\rho D(v,\alpha)\sqrt\rho\,|m\rangle_V\nonumber\\
    &\;=\; \sum_{m,n}{}_V\langle m|D(v,\nu)|n\rangle_V\langle n|\sqrt\rho D(v,\alpha)\sqrt\rho\,|m\rangle_V\nonumber\\
    &\;=\; {\rm tr}\bigl[D(v,\nu)\sqrt\rho D(v,\alpha)\sqrt\rho\,\bigr]\;,
\label{PhiUV}
\end{align}
where in the third step, we use
\begin{equation}
{}_U\langle n|D(u,-\nu^*)|m\rangle_U \;=\;
{}_U\langle n|D(u,-\nu)|m\rangle_U^* \;=\;
{}_U\langle m|D(u,\nu)|n\rangle_U \;=\;
{}_V\langle m|D(v,\nu)|n\rangle_V\;.
\label{cnumconvert}
\end{equation}
Inserting Eq.~(\ref{PhiUV}) into the integral~(\ref{jointPhi}) gives us a form that involves
only system operators:
\begin{align}
\int\frac{\dtwo\nu}{\pi}\,|\Phi_{\rho_{UV}}(\nu,\alpha)|^2
    &\;=\; \int\frac{\dtwo\nu}{\pi}\,\Bigl|{\rm tr}\bigl[D(v,\nu)\sqrt\rho D(v,\alpha)\sqrt\rho\,\bigr]\Bigr|^2 \nonumber\\
    &\;=\; \int\frac{\dtwo\nu}{\pi}\,\sum_{m,n}{}_V\langle m|\sqrt\rho D(v,\alpha)\sqrt\rho D(v,\nu)|m\rangle_V\langle n|D^\dag(v,\nu)\sqrt\rho D^\dag(v,\alpha)\sqrt\rho\,|n\rangle_V \\
    &\;=\; \sum_n{}_V\langle n|\sqrt\rho D(v,\alpha)\sqrt\rho\sqrt\rho D^\dag(v,\alpha)\sqrt\rho\,|n\rangle_V \\
    &\;=\; {\rm tr}\big[\rho D^\dag(v,\alpha)\rho D(v,\alpha)\big] \;.\label{new1}
\end{align}
The simplification in the third step follows from
\begin{equation}
\int\frac{\dtwo\nu}{\pi}\,D(b,\nu)|m\rangle_V\langle n|D^\dag(b,\nu)\;=\;\delta_{mn}I\;,
\label{schur}
\end{equation}
which is a consequence of Schur's Lemma for the (Weyl-Heisenberg)
group of displacement operators, but which can also be derived
directly, for example, from Eq.~(\ref{Dnumbasis}).

The joint Wigner-function integral in Eq.~(\ref{Fentone3}) can now be
obtained by a Fourier transform of all variables in the
characteristic-function integral~(\ref{new1}):
\begin{equation}\label{boldW}
\pi\!\int \dtwo\mu\,W_{\rho_{UV}}(\mu^*,\beta)W_{\rho_{UV}}(\mu^*,\nu)
    \;=\; \frac{1}{\pi^2}\,{\rm tr}\bigl[\rho\tilde D(v,\beta)\rho\tilde D(v,\nu)\bigr]
    \;\equiv\; \bm{W}_\rho(\beta,\nu)\;.
\end{equation}
Here $\tilde D(v,\nu)$ is the Fourier transform of the displacement
operator, as in Eq.~(\ref{tildeD}), and we define the two-variable
function $\bm{W}_\rho(\beta,\nu)$.  For a pure state $\rho$,
$\bm{W}_\rho(\beta,\nu)=W_\rho(\beta)W_\rho(\nu)$, and for mixed
states, $\bm{W}_\rho(\beta,\nu)$ is what replaces the product of
Wigner functions in the third form~(\ref{Fav3}) of the teleportation
fidelity.  The Fourier transform of $\bm{W}_\rho(\beta,\nu)$ is
\begin{equation}\label{boldPhi}
\bm{\Phi}_\rho(\mu,\alpha)
\;=\; \int\dtwo\beta\,\dtwo\nu\,\bm{W}_\rho(\beta,\nu)D^*(\beta,\mu)D(\nu,\alpha)
\;=\; {\rm tr}\big[\rho D^\dag(v,\mu)\rho D(v,\alpha)\big]\;,
\end{equation}
For a pure state $\rho$, we have
$\bm{\Phi}_\rho(\mu,\alpha)=\Phi_\rho^*(\mu)\Phi_\rho(\alpha)$.

We can now summarize our results by rewriting the third and fourth
lines in the entanglement fidelity [Eqs.~(\ref{Fentone3}) and
Eqs.~(\ref{Fentone4})], the lines that tell us about fine-scale
phase-space structure, in terms of the new bold-face functions:
\begin{subequations}
\begin{align}
\Fbent_\rho
    &\;=\;\pi\!\int \dtwo\beta\,\dtwo\nu\,P(\beta-\nu)\bm{W}_\rho(\beta,\nu)\label{Fenttwo3}\\
    &\;=\; \int \dtwo\mu\,\tilde P(\mu)\,\bm{\Phi}_\rho(\mu,\mu)\label{Fenttwo4}\;.
\end{align}
\end{subequations}
These forms can be specialized to the case of squeezed-state
teleportation by inserting the expressions for $P$ and $\tilde P$
from Eqs.~(\ref{Pnu}) and (\ref{tildeP}).

When we repeat the steps of Sec.~\ref{sec3} to find the first
derivative of the entanglement fidelity, the sequence of
Eq.~(\ref{Fderiv1b}) is unchanged, because it uses the upper Fourier
pair in Eq.~(\ref{Fentone}), but the sequence of
Eq.~(\ref{Fderiv1a}), which uses the lower Fourier pair, must be
modified to use the bold-face functions.  The resulting expressions
for the derivative,
\begin{equation}
\left.\frac{d\Fbent_\rho}{dt}\right|_{t=0}
    \;=\; -\frac{1}{2}\left({\Delta x_V}^2+{\Delta p_V}^2\right)
    \;=\; -\frac{\pi}{2}\int\dtwo\nu\,\left.
    \frac{\partial^{\hspace{0.1em}2}\bm{W}_\rho(\nu,\alpha)}{\partial\nu^*\,\partial\alpha}\right|_{\alpha=\nu}\;,
\end{equation}
give rise to the mixed-state generalizations of our measures for
small- and large-scale phase space structure:
\begin{equation}
\left(-2\left.\frac{d\Fbent_\rho}{dt}\right|_{t=0}\right)^{1/2} \;=\;
\frac{1}{\ell_c} \;=\; \frac{L_c}{2} \;=\; \sqrt{{\Delta x_V}^2+{\Delta p_V}^2}
\;=\;\left(\pi\int\dtwo\nu\,\left.
    \frac{\partial^{\hspace{0.1em}2}\bm{W}_\rho(\nu,\alpha)}{\partial\nu^*\,\partial\alpha}\right|_{\alpha=\nu}\right)^{1/2}\;.
\end{equation}

It is instructive to see how this works out for a thermal state,
\begin{equation}
\rho=(1-e^{-\lambda})e^{-\lambda b^\dagger b}\;=\;
\int\frac{\dtwo\beta}{\pi\bar n}\,e^{-|\beta|^2/\bar n}|\beta\rangle\langle\beta|\;,
\end{equation}
where $\lambda$ is the dimensionless inverse temperature and $\bar
n=(e^\lambda-1)^{-1}$ is the mean number of quanta.  The second form
is the standard $P$-function representation of a thermal
state~\cite{Cahill1969}.  Inserting the $P$-function representation
into the expression~(\ref{boldPhi}) gives
\begin{equation}
\bm{\Phi}(\mu,\alpha) \;=\;
    \frac{1}{2\bar n+1}\exp\!\left(-\frac{2\bar n^2+2\bar n+1}{2(2\bar n+1)}\bigl(|\mu|^2+|\alpha|^2\bigr)
    +\frac{\bar n(\bar n+1)}{2\bar n+1}(\mu\alpha^*+\mu^*\alpha)\right)
\end{equation}
and thus
\begin{equation}
\bm{\Phi}(\mu,\mu) \;=\; \frac{1}{2\bar n+1}e^{-|\mu|^2/(2\bar n+1)}\;.
\end{equation}
The resulting average entanglement fidelity from Eq.~(\ref{Fenttwo4})
is
\begin{equation}
\Fbent_\rho \;=\; \frac{1}{1+(2\bar n+1)t/2}\;.
\end{equation}
This gives
\begin{equation}
\frac{1}{\ell_c}=\frac{L_c}{2}=\sqrt{2\bar n+1}\;,
\end{equation}
which is consistent with the variances of $x$ and $p$, i.e., $(\Delta
x)^2+(\Delta p)^2=2\bar n+1$.

\section{Conclusion}
\label{sec6}

In this paper we examined the relationships among the output fidelity
of continuous-variable teleportation protocols, sub-Planck structure
in the Wigner function of the teleported state, and the large-scale
extent of that Wigner function.   For pure states, these
relationships are made mathematically precise in
Eqs.~(\ref{Fderiv1a}) and (\ref{Fderiv1b}), which lead us to define
measures of small- and large-scale structure for the Wigner function
in Eqs.~(\ref{ellc}) and (\ref{Lc}).  Consideration of several
example states in Sec.~\ref{sec4} illuminates these relationships and
builds confidence that the measures of small- and large-scale
structure we define are quite reliable measures of phase-space
properties of the Wigner function.

The generalization of these results to mixed states in
Sec.~\ref{sec6} leads to a pair of new functions, generalizations of
the Wigner and characteristic functions, which capture the fine-scale
phase-space structure in any purification of the mixed state.  These
new functions might prove useful in other studies of phase-space
properties of mixed states.

\begin{acknowledgments}
This work was supported in part by Office of Naval Research Grant
No.~N00014-07-1-0304 and by National Science Foundation Grant
No.~PHY-0653596.  AJS acknowledges support from the Australian
Research Council and the State of Queensland.
\end{acknowledgments}

\end{document}